\documentclass[a4paper,11pt]{article}
\usepackage{jheppub} 
\usepackage{ccaption}
\usepackage{float}
\usepackage{hyperref}

\title{\boldmath A Differential Form Description of Partial Entanglement Entropy: Testing a Killing Vector Construction in Covariant Phase Space}

\author{Chuanjia Zhu$^{1,2}$, Dong-Hui Du$^{3}$, Wen-Cong Gan$^{4}$ and Fu-Wen Shu$^{1,2}$}
\affiliation{	$^{1}$Department of Physics, Nanchang University, Nanchang, 330031, China\\
	$^{2}$Center for Relativistic Astrophysics and High Energy Physics, Nanchang University, Nanchang, 330031, China\\ $^{3}$Department of Physics, School of Science, East China University of Technology, Nanchang, 330013, China\\
    $^{4}$School of Physics, Jiangxi Normal University, Nanchang, 330022, China}

\emailAdd{chuanjiazhu@email.ncu.edu.cn, donghuidu@ecut.edu.cn, ganwencong@jxnu.edu.cn, shufuwen@ncu.edu.cn}

\abstract{The bit thread formulation provides a geometric description of holographic entanglement entropy, while partial entanglement entropy (PEE) and PEE threads resolve this structure with respect to individual boundary points. Motivated by the fact that PEE thread flows can be superposed to reconstruct conventional bit thread flows, we introduce a differential form description of PEE thread flow. For an interval in the vacuum AdS$_3$/CFT$_2$ setup, we show that the flux of the resulting form reproduces the known entanglement contour, providing a consistency check of the proposed description. We then investigate whether the PEE form can be related, possibly up to an exact form improvement, to a current constructed from the covariant phase space (CPS) formalism. As a first test, we consider exact Killing vectors in Rindler-AdS$_3$. The Rindler parameter $a$ parametrizes a one-parameter family of backgrounds. At each value of $a$, the Killing vector of the corresponding background is substituted into the Iyer–Wald surface charge form, and the full $a$-dependence of the resulting form is retained. We define a finite candidate current by integrating this surface charge form over $a$. For the reflection-symmetric PEE flow sourced at the central boundary point $r_0=0$, we find that no choice of the Killing parameters reproduces both components of the known PEE flow after transforming to Poincaré-AdS$_3$. This result shows that the direct Killing vector CPS construction considered here is not sufficient to reproduce the central PEE flow. Possible extensions include an exact form improvement, a different CPS current, or a more general choice of generator.}

\begin{document}
 \maketitle

\section{Introduction}
\label{sec:intro}
In the context of the AdS/CFT correspondence \cite{Maldacena:1997re}, the Ryu-Takayanagi (RT) formula proposed that the entanglement entropy (EE) for an arbitrary spatial region $A$ in the boundary CFT is given by the area of a minimal surface in the AdS bulk \cite{Ryu:2006bv}, that is
\begin{equation}
\label{20}S(A)=\frac{\text{Area}(m_A)}{4G_N},
\end{equation}
where $m_A$ represents the minimal surface that is homologous to $A$. In the case of a time-dependent and non-static spacetime, a covariant proposal of entanglement entropy was established in \cite{Hubeny:2007xt}, with subsequent developments of its quantum-corrected counterpart \cite{Faulkner:2013ana,Engelhardt:2014gca}. Complementing these advances, holographic emergence of bulk spacetime geometry from CFT data—particularly metric reconstruction from boundary observables—represents a consequential research direction leveraging the RT formula. The incomplete list of related works refer to \cite{Hammersley:2006cp,Bilson:2008ab,Bilson:2010ff,Hammersley:2007ab,Czech:2012bh,deHaro:2000vlm,Gan:2016vuw,Gan:2016vjt}.
\par Subsequently, Freedman and Headrick reformulated the RT formula using the max flow-min cut (MFMC) theorem, in a way that does not make reference to the minimal surface \cite{Freedman:2016zud}. They introduced a notion of ``flow'', defined as a divergenceless norm-bounded vector field on manifolds. Within this framework, the area of the minimal surface $m_A$ is obtained by maximizing flux through the boundary region $A$ by optimizing over all possible divergenceless vector fields $V$ on the manifold, where the norm of $V$ is upper-bounded as $|V|\leq\frac{1}{4G_N}$. For an optimized flow, the vector $V$ should be normal to the minimal surface $m_A$ with the norm saturating the bound $1/4G_N$. Hence, the minimal surface $m_A$ plays the role of a bottleneck that constrains the flow configurations on the manifold. The
integral curves of the optimized flow configuration are called bit threads, which emanate from the boundary region $A$ and end on its complement. We should point out that the configuration of bit threads does not lead to unique usage by the prescription. \par
Subsequent developments have established bit thread configurations in pure AdS spacetime using bulk geodesics \cite{Agon:2018lwq}. Similar to the RT formula, bit threads have been extended to a quantum-corrected version and a covariant version \cite{Chen:2018ywy,Agon:2021tia,Rolph:2021hgz,Headrick:2022nbe}. The bit thread formulation of the holographic entanglement entropy in higher curvature gravity has also been constructed \cite{Harper:2018sdd}. Furthermore, these constructions have been employed to explore the first law of entanglement in relation to Einstein’s equations \cite{Agon:2020mvu} and so on. For other references and recent progress on bit threads, please refer to \cite{Du:2019emy,Du:2024xoz,Du:2019vwh,Chen:2018ody,Lin:2025jjh,Das:2025fav}.\par
In order to describe how many degrees of freedom at each site contribute to the entanglement between two separate parts, it is useful to invoke the notion of the entanglement contour. The entanglement contour was first proposed in \cite{Vidal:2014aal}, and the relation between the entanglement contour and the bit thread configuration was first pointed out in a talk by Erik Tonni \cite{Tonni}. More explicitly, the entanglement contour could be regarded as a special partial entanglement entropy (PEE) \cite{Kudler-Flam:2019oru,Wen:2018whg,Han:2019scu,Wen:2019iyq,Wen:2020ech,Han:2021ycp} which is a measure of the correlation between two spacelike separated regions $A$ and $B$. Recently, a scheme of geometrizing the PEE in the context
of AdS/CFT has been constructed \cite{Lin:2023rxc}. Given a point $x$, we can geometrize the two-point PEEs \cite{Wen:2020ech} between $x$ and any other points in terms of the bulk geodesics connecting these two parts. We refer to these geodesics as the PEE threads, which can be naturally regarded as the integral curves of a divergenceless vector field $V_x^a$ (or PEE thread flow) \cite{Lin:2023rxc}. Crucially, bit threads emerge from PEE threads through integration of this flow over the boundary region $A$, demonstrating how non-intrinsic bit threads originate from intrinsic PEE structures \cite{Lin:2023rxc}.\par
Based on the general methods originally developed in \cite{Lashkari:2013koa}, the perturbative thread configuration has been constructed in \cite{Agon:2020mvu}. These constructions, however, encode bulk metric information in a nonlocal manner. To address this limitation, the authors reformulated the bit thread framework using differential forms. In \cite{Das:2025fav}, an expression for the bit threads vector field generated from covariant phase space(CPS) has been constructed. Given that bit threads emerge from PEE threads, a natural extension involves formulating a differential framework for PEE threads and how to construct PEE thread vectors from CPS. This constitutes the primary objective of our work.\par
In this work, we first introduce a differential form description of PEE thread flows by taking the Hodge dual of the corresponding divergenceless vector field on a spatial bulk slice. For an interval in vacuum AdS$_3$/CFT$_2$, we show that the flux of the resulting form reproduces the known entanglement contour, providing a consistency check of the proposed formulation. Motivated by covariant phase space constructions of ordinary bit threads, we then investigate whether the PEE form can be related, possibly up to an exact form improvement, to a current constructed from the CPS surface charge form. As a first test, we consider exact Killing vectors in Rindler-AdS$_3$. We use the Rindler parameter $a$ to parametrize a one-parameter family of backgrounds and write the metric variation as 
\begin{equation}
\delta \hat{g}_{\mu\nu}(a)=\frac{\partial \hat{g}_{\mu\nu}}{\partial a}\delta a
\end{equation}
At each value of $a$, the Killing vector $\tilde{\xi}(a)$ of the corresponding background is substituted into the Iyer–Wald surface charge form. The resulting form therefore contains the $a$-dependence of both the background and the Killing solution. We define a finite candidate current by integrating this surface charge form over $a$. Restricting the PEE source to the center of the boundary interval, $r_0=0$, we transform the associated candidate flow to Poincaré-AdS$_3$ and compare it with the known reflection-symmetric PEE flow. We find that no choice of the Killing parameters reproduces both components of the known PEE flow. This shows that the direct Killing vector CPS construction considered here is not sufficient, while more general CPS realizations remain possible.
\par
The remainder of this paper is organized as follows. In Section \ref{section 1}, we briefly review the bit thread formulation, partial entanglement entropy, and the construction of PEE threads. Further details about the differential form of bit threads can be found in Appendix \ref{app}. In Section \ref{sec 2}, we review the differential form description of bit threads and introduce the corresponding differential form for PEE flows. We also discuss its linear perturbation and perform a consistency check by recovering the entanglement contour for an interval in AdS$_3$. In Section \ref{sec 4}, we propose a candidate relation between the PEE form and a current constructed from the CPS surface charge form. We then test an exact Killing vector construction in Rindler-AdS3$_3$. At each value of $a$, the Killing vector of the corresponding background is used to evaluate the surface charge form, which is integrated over $a$ to define a finite candidate current. The associated flow is transformed to Poincaré-AdS$_3$ and compared with the known central PEE flow. Section \ref{DC} summarizes our results, discusses the scope of the mismatch found in this calculation, and considers possible extensions of the proposed PEE–CPS relation.\par
\section{A brief review of bit threads and PEE threads}\label{section 1}
\subsection{Bit threads}

The bit-thread formulation has reformulated the RT formula by using the MFMC theorem. According to the formulation, the entanglement entropy of the boundary region $A$ can be obtained by maximizing the flux of a divergenceless vector field $V_A^a$ through any co-dimension 2 surface $\Sigma$ homologous to region $A$, i.e.
\begin{equation}
\label{1}S_A=\text{max}\int_{\Sigma}\mathrm{d}\Sigma\sqrt{g}V_A^an_a,
\end{equation}
where $n_a$ is the normal unit vector on co-dimension 2 surface and $g$ is the determinant of the induce metric on the surface. The divergenceless condition $\nabla_aV_A^a=0$ means that the bit threads cannot begin, end, split, or join in the bulk. Therefore, each thread begins and ends only on the boundary.\par 
Except for the requirement of divergencelessness, the vector field should satisfy the inequality:
\begin{equation}
|V^a_A|\leq\frac{1}{4G_N},
\end{equation}
which saturates on the RT surface $m_A$, simultaneously, the vector should be normal to the RT surface. Obviously, the minimal surface is a bottleneck such that any flow from region $A$ is bounded by this bottleneck. In other words, the minimal surface is the place where the bit threads happen to be most tightly packed together. There is freedom to choose the flow beside the bottleneck; thus, whereas the minimal surface is generically unique, the flux-maximizer generically enjoys an enormous degeneracy, hence the configuration of bit threads has no physical meaning. The integral curves of the flux-maximizing flow are called bit threads.

\subsection{PEE and PEE threads}

Partial entanglement entropy (PEE) or entanglement contour \cite{Kudler-Flam:2019oru,Wen:2018whg,Han:2019scu,Wen:2019iyq,Wen:2020ech,Han:2021ycp} is a quasi-local measure of entanglement; in other words, PEE or entanglement contour captures the contribution of each local degree of freedom in a given subsystem to the total entanglement. If we know the entanglement contour function $s_A(x)$ of a boundary region $A$, the entanglement entropy $S_A$ is just the collection of contributions from all degrees of freedom inside region $A$, that is
\begin{equation}
\label{16}S_A=\int_A s_A(x)\mathrm{d}\sigma_{x}.
\end{equation}
where $\mathrm{d}\sigma_x$ is an infinitesimal area element located at $x$.
In other words, given a subregion $A_i$ in $A$, the partial entanglement entropy $s_A(A_i)$ can be written as
\begin{equation}
s_A(A_i)=\int_{A_i}s_A(x)d\sigma_x.
\end{equation}\par
To endow bit threads with a more obvious physical meaning, it may be more realistic to take the bit thread configuration as some emergent concepts from a certain intrinsic structure of the state. It has been proposed in \cite{Lin:2023rxc} that we can geometrize the PEE to get the PEE threads, and the most important point is that the bit threads can be reconstructed by PEE threads. To understand the point, we should note that any PEE $\mathcal{I}(A,B)$ can be evaluated by the integration (or summation for discrete systems) of a certain class of two-point PEEs \cite{Wen:2020ech}:
\begin{equation}
\label{2}\mathcal{I}(A,B)=\int_A \mathrm{d}\sigma_x\int_B \mathrm{d}\sigma_y\mathcal{I}(x,y),
\end{equation}
Most importantly, using formula \eqref{2}, the entanglement entropy of region $A$ between its complementary $\bar{A}$ can be rewritten as 
\begin{equation}
\label{17}S_A=\int_A \mathrm{d}\sigma_x\int_{\bar{A}} \mathrm{d}\sigma_y\mathcal{I}(x,y)=\mathcal{I}(A,\bar{A}).
\end{equation}
The above formula means that we can recognize the entanglement entropy as a collection of all the two-point PEEs, between $A$ and $\bar{A}$.\par
\begin{figure}[ht] 
   \centering
    \includegraphics[width=0.9 \textwidth]{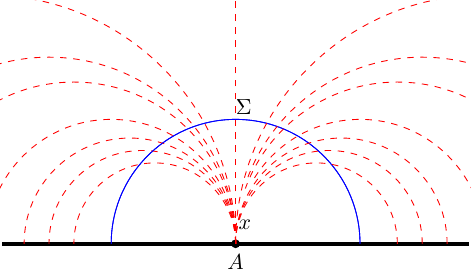}\quad
 \caption{It shows the \textit{PEE threads} emanating from a boundary point $x$, the flow $V_x^a$ is tangent to the PEE threads (the red dashed curves). $\Sigma$ (the blue curve) denotes a co-dimension two surface that is homologous to the boundary region $A$. The norm of $V_x^a$ is determined by requiring the flux of $V_x^a$ on $\Sigma$ to be equal to the entanglement contour $s_A(x)$.
\label{Fig1}}
\end{figure}

Since the entanglement contour can be considered as a special PEE, from the above formulas \eqref{17} and \eqref{16}, we can get 
\begin{equation}
s_A(x)=\mathcal{I}(x,\bar{A})=\int_{\bar{A}}\mathrm{d}\sigma_y\mathcal{I}(x,y).
\end{equation}
Then, we can geometrize the two-point PEEs in
terms of their corresponding bulk geodesics and name these geodesics bundles as the PEE threads. For any boundary point $x$, the PEE threads emanating from it can be understood as the integral curves of a divergenceless vector field $V^a_x$. We have used the subscript $x$ to distinguish the bit threads vector $V_A^a$ from the PEE threads vector, and more importantly, the subscript $x$ represents the PEE threads emanating from boundary points $x$.  The vector field is called the PEE thread flow vector field (see Fig. \ref{Fig1}). To make good use of the physical picture, we can reasonably require that the contribution of site $x$ to $S_A$ or the value of $s_A(x)$ is captured by the number of PEE threads connecting $x$ and $\bar{A}$ \cite{Lin:2023rxc}, that is
\begin{equation}
\label{13}s_A(x)=\int_{\bar{A}}\mathrm{d}\sigma_y\mathcal{I}(x,y)=\int_{\Sigma}\mathrm{d}\Sigma\sqrt{g}V_x^an_a,
\end{equation}
where $n_a$ is the unit vector normal to $\Sigma$.
\par
\begin{figure}[ht] 
\centering
\includegraphics[width=0.9 \textwidth]{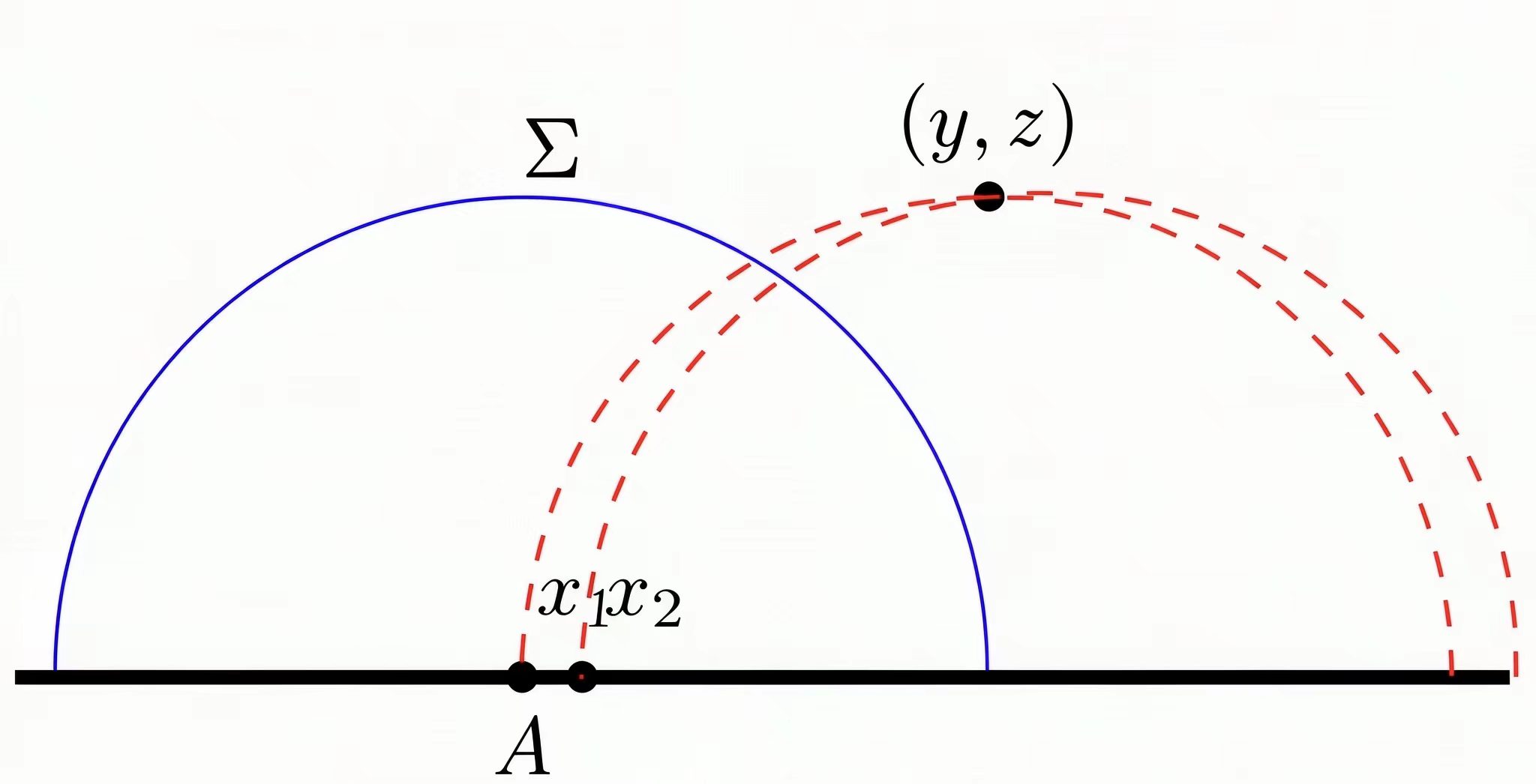}\quad
\caption{For any bulk points $(y,z)$, there are infinitely many PEE threads emanating from one boundary point that penetrate it ( $x_1,x_2$ as indicated in the figure). To get bit threads from PEE threads we should integrate all such points in the boundary region $A$.
\label{Fig2}}
\end{figure}

 For all points $x$ on the boundary, there will be PEE threads emanating from it. Therefore, the entanglement entropy is obtained by collecting all the PEE threads coming out of $A$. As a consequence, we should superpose all the PEE thread flows $V_x^a$ with $x$ inside $A$. Fortunately, the PEE threads that emanate and terminate inside $A$ will not contribute to the final results. Therefore, there is no need to distinguish between the inner and outer threads, and the bit thread flow $V_A^a$ can be achieved by integrating the whole PEE thread flow $V_x^a$ with $x\in A$ \cite{Lin:2023rxc}, i.e. 
\begin{equation}
\label{3}V_A^a=\int_A \mathrm{d}^{d-1}xV_x^a.
\end{equation}
It is important to note that the integral of the variable $x$ here operates on the subscript of $V^a_x$. Since we should collect all PEE threads emanating from all points of the boundary region $A$ that penetrate the bulk point $(y,z)$. So, there is another spatial variable contained in $V^a_x$, we can rewrite the expression as follows to prevent ambiguity, i.e.
\begin{equation}
\label{26}V_A^a(y,z)=\int_A \mathrm{d}^{d-1}xV_{x}^a(y,z).
\end{equation}
 where $x$ represents the boundary integration variable, $(y,z)$ is the variable of one point in the bulk (see Fig. 
  \ref{Fig2}). 
The above formula means that by collecting all PEE thread vector fields emanating from all boundary points $x$, we obtain the bit thread vector field. Hence, the non-intrinsic bit threads are emergent from the intrinsic PEE threads.

\section{Differential form of bit threads and PEE threads}\label{sec 2}
In this section, we first review the differential form of bit threads and linear perturbation. Then we propose a scheme to give the differential form of PEE threads, and we have also verified our differential form in AdS$_3$/CFT$_2$ for the sphere region, i.e., using the differential form we successfully extract the entanglement contour for the sphere region in AdS$_{3}$/CFT$_2$.
\subsection{A brief review of differential form of bit threads and linear perturbation}
In this subsection, for the convenience of the reader and the subsequent narrative, we will first briefly review the simple realizations of perturbative bit threads and the differential form of bit threads. \par
In the context of pure AdS$_{d+1}$, the geometry of a constant $t$ slice $\Sigma$ is given by
\begin{equation}
ds^2=\frac{1}{z^2}(\mathrm{d}r^2+r^2\mathrm{d}\Omega^2_{d-2}+\mathrm{d}z^2),
\end{equation}
Now, we consider a linear perturbation of the bulk metric, i.e., $g^{\lambda}_{\mu\nu}=g_{\mu\nu}+\lambda\delta g_{\mu\nu}+\mathcal{O}(\lambda^2)$. In Fefferman-Graham coordinates, the metric perturbation takes the form $\delta g_{ij}=z^{d-2}H_{ij}$ with $\delta g_{zz}=\delta g_{zi}=0$. Naturally, there will be a linear perturbation about the bit threads vector field, $V^a_{A,\lambda}=V_A^a+\lambda\delta V_A^a+\mathcal{O}(\lambda^2)$. Hence, the PEE threads vector field also gets a linear perturbation as
\begin{equation}
\label{4}V^a_{x,\lambda}=V^a_x+\lambda\delta V^a_x+\mathcal{O}(\lambda^2).
\end{equation}
We assume that the perturbed PEE vector field still constructs the bit-thread vector field, hence by using formulas \eqref{3} and \eqref{4}, we can arrive at
\begin{equation}
V_{A,\lambda}^a=\int_A \mathrm{d}^{d-1}xV^a_{x,\lambda}=\int_A \mathrm{d}^{d-1}xV^a_x+\lambda\int_A \mathrm{d}^{d-1}x\delta V^a_x.
\end{equation}
Here we only consider the linear order. Obviously, the first term on the right of the second equality is the bit threads before perturbation. As a consequence, we get
\begin{equation}
\label{5}\delta V^a_A=\int_A \mathrm{d}^{d-1}x\delta V_x^a.
\end{equation}
This equation means that the integral of all the perturbed PEE-thread vector fields emanating from the boundary points $x$ will give the perturbed bit threads vector field.\par
As discussed in \cite{Agon:2020mvu}, we can rewrite formula \eqref{20} in differential form version as
\begin{equation}
\label{25}S_A=\mathop{\text{max}}_{\substack{\boldsymbol{w}\in\boldsymbol{W}}}\int_A\,\boldsymbol{w},
\end{equation}
where $\boldsymbol{W}$ is the set of closed forms obeying the bound
\begin{equation}
\frac{1}{(d-1)!}g^{a_1b_1}\dots g^{a_{d-1}b_{d-1}}w_{a_1\dots a_{d-1}}w_{b_1\dots b_{d-1}}\le\Big(\frac{1}{4G_N}\Big)^2
\end{equation}
and $\boldsymbol{w}$ is a $(d-1)$ differential form related to divergenceless vector fields $V^a$, i.e.
\begin{equation}
\label{29}{V^a_A=g^{ab}(\boldsymbol{\star{w}})_b,}
\end{equation}
where ‘$\star$’ is the Hodge dual defined on a Cauchy surface $\Sigma$, defined via
\begin{equation}
    (\boldsymbol{\star{w}}_b)=\frac{1}{(d-1)!}\sqrt{g}w^{a_1\dots a_{d-1}}\varepsilon_{a_1\dots a_{d-1}b},
\end{equation}
and $\varepsilon_{a_1\dots a_{d-1}}$ represents the totally antisymmetric Levi-Civita symbol, with sign convention $\varepsilon_{i_1\dots i_{d-1}z}=1$.
More details on the derivation and $\boldsymbol{w}$ can be found in Appendix \ref{app}.\par
Since we associate a $(d-1)$ differential form $\boldsymbol{w}$ to the divergenceless vector field of bit threads, and take advantage of the nice properties of $ \boldsymbol{\mathcal{\chi}}$ \cite{Faulkner:2013ica} defined via the Iyer-Wald formalism \cite{Iyer:1994ys,Iyer:1995kg}, we could have \cite{Agon:2020mvu}
\begin{equation}
\label{30}\boldsymbol{\mathcal{\chi}}=\delta \boldsymbol{w}.
\end{equation}
More importantly, it can be related to $\delta\boldsymbol{w}$ via its Hodge dual by introducing a new vector field $\delta 
 V^a_{\Phi,A}$ \cite{Agon:2020mvu}, defined by
\begin{equation} 
\label{15}\delta V_{\Phi,A}^a=g^{ab}(\star\delta \boldsymbol{w})_b,
\end{equation}
where $\delta V_{\Phi,A}^a=\delta V^a_A+\frac{1}{2}g^{bc}\delta g_{bc}V^a_A$.
And using this proposal, we can write the perturbed bit threads vector field as follows \cite{Agon:2020mvu}:
\begin{equation}
\begin{aligned}
\label{6}&\delta V^{\mu}_A=\frac{z^{d+1}}{16\pi G_N}\left\{\left[\left(\frac{2\pi z}{R}+\frac{d}{z}\xi^t+\xi^t\partial^z\right)H^i_i-\frac{8\pi G_N}{z}V_A^zH^i_i\right]\partial_z-\left(\frac{8\pi G_N}{z}V^i_A\right)H^j_j\partial_i\right.\\
&\left. \qquad+\left[\left(\frac{2\pi(x^i-x^i_0)}{R}+\xi^t\partial^i\right)H^j_j-\left(\frac{2\pi(x^j-x^j_0)}{R}+\xi^t\partial^j\right)H^i_j\right]\partial_i\right\},
\end{aligned}
\end{equation}
where $\xi^t$ is a Killing vector \cite{Faulkner:2013ica}
\begin{equation}
\xi=-\frac{2\pi}{R}(t-t_0)[z\partial_z+(x^i-x^i_0)\partial_i]+\frac{\pi}{R}[R^2-z^2-(t-t_0)^2-(\vec{x}-\vec{x_0})^2]\partial_t.
\end{equation}
Since, as mentioned in \cite{Agon:2020mvu}, in the presence of a metric, the set of $(d-1)$ forms $\delta \boldsymbol{w}$ defines a set of covector fields $\delta w_a(R,\vec{x_0},z,\vec{x})$ i.e. by formula \eqref{15} and using these covectors, we can reconstruct the bulk perturbed metric $\delta g_{ij}$ locality. \par
{\subsection{Differential form of PEE threads}}
It is natural to notice that PEE forms may provide a point-resolved decomposition of the perturbative data relevant for metric reconstruction. It should be noted that, similar to bit threads, we also need a PEE thread construction that does not make explicit use of the metric.
To gain benefit from the canonical form, we want to get a differential form of PEE threads that we can use to reconstruct the bulk perturbed metric. But how do we find the differential language of PEE threads? 
The flux of the PEE thread flow $V_x^{\mu}$ on any co-dimension two bulk hypersurface $\Sigma$ homologous to a boundary region $A$ should satisfy the following
\begin{equation}
s_A(x)=\int_{\Sigma}\boldsymbol{\tilde{\epsilon}}(V_x^an_a),
\label{22}\end{equation}
where $V_x^a$ represents the PEE threads flow vector field, which is divergenceless. $\boldsymbol{\tilde{\epsilon}}$ is volume element on $\Sigma$ and $n_a$ is the unit vector normal to $\Sigma$. Actually, if we choose hypersurface $\Sigma=A$, using our notion \eqref{26} we can write \eqref{22} as 
\begin{equation}
s_A(x)=\int_{\Sigma=A}\boldsymbol{\tilde{\epsilon}}(V^a_{x}(y,z=0)n_a)
\end{equation}
For convenience, in the following content, we will omit variables in parentheses, which should be kept in mind. 
Notice that the integrand is similar to formula \eqref{1}, just replace $V_A^a$ with $V_x^a$. So we could formally associate a differential form $\boldsymbol{\tilde{w}}$ with a divergenceless vector field $V_x^a$. We assume that $\boldsymbol{\tilde{w}}$ is given by
\begin{equation}
\label{27}{V_x^a}=g^{ab}(\star\boldsymbol{\tilde{w}})_b,
\end{equation}
which is similar to formula \eqref{29}, and the $(d-1)$ form $\boldsymbol{\tilde{w}}$ is given by:
\begin{equation}\label{eq3.16}
\boldsymbol{\tilde{w}}=\frac{1}{(d-1)!}\epsilon_{\mu_1\dots\mu_{d-1}b}V_x^b \mathrm{d}x^{\mu_1}\wedge\dots\ \wedge \mathrm{d}x^{\mu_{d-1}}.
\end{equation}
where $\epsilon_{\mu_1\dots\mu_{d-1}b}$ are the components of bulk volume elements, which satisfy $\epsilon_{i_1\dots i_{d-1}z}=\sqrt{g}$. Taking the exterior derivative of the above equation also led to:
\begin{equation}
\label{21}\mathrm{d}\boldsymbol{\tilde{w}}=(\nabla_aV_x^a)\boldsymbol{\epsilon}.
\end{equation}
Next, by considering Gauss's theorem applied to the divergenceless vector field $V_x^a$ in a bulk region $N$ with $\partial N=A\cup(-\Sigma)$, we have
\begin{equation}
\label{24}\int_N(\nabla_aV_{x}^a)\boldsymbol{\epsilon}=\int_{\partial N}(V_{x}^an_a)\boldsymbol{\tilde{\epsilon}}=\int_A(n_aV_{x}^a)\boldsymbol{\tilde{\epsilon}}-\int_{\Sigma}(n_aV_{x}^a)\boldsymbol{\tilde{\epsilon}}=0.
\end{equation}
where $n^a$ is local unit normal, and $\boldsymbol{\epsilon}$, $\tilde{\boldsymbol{\epsilon}}$ is the bulk and boundary volume element respectively.
And using \eqref{21} we have
\begin{equation}
\label{23}\int_N\nabla_aV_x^a\boldsymbol{\epsilon}=\int_N\mathrm{d}\boldsymbol{\tilde{w}}=\int_{\partial N}\boldsymbol{\tilde{w}}=\int_A\boldsymbol{\tilde{w}}-\int_{\Sigma}\boldsymbol{\tilde{w}}=0
\end{equation}
Combining the above equations \eqref{22}, \eqref{24} and \eqref{23}, we have
\begin{equation}
\int_A\boldsymbol{\tilde{w}}=\int_{\Sigma}\boldsymbol{\tilde{w}}=s_A(x)
\end{equation}

The above formula means that the integral of the PEE differential on any codimension-two bulk hypersurface $\Sigma$ homologous to a boundary region $A$ should recover the entanglement contour of the boundary region $A$. 
Similarly, we can define a new vector field $\delta V^a_{\Phi,x}$ 
\begin{equation}
\label{18}\delta V^a_{\Phi,x}=\delta V^a_x+\frac{1}{2}g^{bc}\delta g_{bc} V^a_x,
\end{equation}
and it is related to $\delta \tilde{\boldsymbol{w}}$ via its Hodge dual
\begin{equation}
\label{19}\delta V^a_{\Phi,x}=g^{ab}(\star\delta\tilde{\boldsymbol{w}})_b,
\end{equation}
according to the above formulas  \eqref{27}, \eqref{18} and \eqref{19}, we have 
\begin{equation}
\delta V^a_{x}=g^{ab}(\star\delta\tilde{\boldsymbol{w}})_b-\frac{1}{2}g^{dc}\delta g_{dc}g^{ae}(\star{\tilde{\boldsymbol{w}}})_e.
\end{equation}
Notice that from formula \eqref{3}, we have
\begin{equation}
\label{8}\delta V^a_A=\int_A \mathrm{d}^{d-1}x[g^{ab}(\star\delta\tilde{\boldsymbol{w}})_b-\frac{1}{2}g^{dc}\delta g_{dc}g^{ae}(\star{\tilde{\boldsymbol{w}}})_e].
\end{equation}\par
In order to guarantee the formula works well and correctly, we must verify our proposal. In the next section, we show that using our proposal, we can obtain the entanglement contour function for static spherical regions in states dual to Poincaré AdS$_{3}$. This means that our proposal is correct to at least some extent.

\

\subsubsection{A consistency check using the known AdS$_3$ PEE flow}
In order to check out our proposal for the PEE threads differential form, we first consider the situation of an interval region with radius $R$ in AdS$_3$. We will use our formula for the PEE differential form to reconstruct the entanglement contour of the interval region on the boundary. In AdS$_3$, the Poincaré metric on a static time slice is given by
\begin{equation}
ds^2=\frac{1}{z^2}(\mathrm{d}r^2+\mathrm{d}z^2),
\end{equation}
where we set AdS radius $L$ equal to $1$. According to the AdS/CFT correspondence, gravitational theories on AdS$_3$ space of radius 1 are dual to 2$d$ CFT with the central charge \cite{Brown:1986nw}:
\begin{equation}
c=\frac{3}{2G_N^{(3)}}.
\end{equation}
As discussed in \cite{Lin:2023rxc}, for any arbitrary bulk point $(r,z)$, the PEE threads vector field at $r_0$ is expressed by
\begin{equation}
\label{37}V_{r_0}^{\mu}=\frac{1}{4G_N}\frac{2z^2(r-r_0)}{((r-r_0)^2+z^2)^2}\left(z,\frac{z^2-(r-r_0)^2}{2(r-r_0)}\right).
\end{equation}
Using our definition of the PEE differential form Eq.\eqref{eq3.16}, we can get
\begin{equation}
\begin{aligned}
\boldsymbol{\tilde{w}_{r_0}}
&=\boldsymbol{\epsilon_{rz}}V_{r_0}^z(dr)+\boldsymbol{\epsilon_{zr}}V_{r_0}^r(dz)\\
&=\sqrt{g}(V_{r_0}^z(dr)-V_{r_0}^r(dz))\\
&=\frac{1}{z^2}(V_{r_0}^z(dr)-V_{r_0}^r(dz)).
\end{aligned}
\end{equation}
Since our integral is on the boundary region, the second term should be dropped, taking the expression of $V_{r_0}^z$ leads to
\begin{equation}
\frac{1}{z^2}\int_{-R}^RV_{r_0}^z \mathrm{d}r=\frac{1}{4G_N}\frac{2R(R^2-r_0^2+z^2)}{((R-r_0)^2+z^2)((R+r_0)^2+z^2))}.
\end{equation}
Since on the boundary, the component of $z$ is equal to 0, i.e., we could drop the term $z$, hence
\begin{equation}
\begin{aligned}
s_A(r_0)
&=\int_A\boldsymbol{\tilde{w}_{r_0}}\\
&=\frac{1}{z^2}\int_{-R}^RV_{r_0}^z \mathrm{d}r\\
&=\frac{1}{4G_N}\frac{2R}{(R^2-r_0^2)}\\
&=\frac{c}{6}\left(\frac{2R}{R^2-r_0^2}\right).
\end{aligned}
\end{equation}
where in the last equal sign we used the definition of the central charge. The result is in perfect agreement with the result in \cite{Kudler-Flam:2019oru}. 
\section{Candidate relation of PEE threads from covariant phase space}\label{sec 4}
In the previous chapter, we proposed a differential form of PEE and used it to derive the entanglement contour. For the sake of greater rigor, we utilize means of CPS formalism to associate with our PEE differential form. Another reason is that since bit threads can be generated from PEE threads and in recent work \cite{Das:2025fav}, a method of bit threads vector field constructed from CPS has been proposed
\begin{equation}
V_{cps}^a=g^{ab}(\star \boldsymbol{j}_{\xi})_b,
\end{equation}
where $\boldsymbol{j}_{\xi}$ is a codimension-2 current and $(d-1)$ form from the CPS and $\xi$ is killing vector field, where $g^{ab}$ is induced metric on constant-time slice. So we could naturally ask about how to construct PEE threads vector field from the covariant phase space? Maybe we also write 
\begin{equation}
\label{31}\boldsymbol{k}_{\xi}=\delta\boldsymbol{j}_{\xi}=\delta\boldsymbol{Q}_{\xi}-\xi\cdot\boldsymbol{\Theta}.
\end{equation}
Equation \eqref{31} is the variational relation underlying the covariant phase space construction of ordinary bit thread flows. In that construction, the standard Iyer–Wald current may have to be supplemented by an additional contribution associated with the ambiguity of the Noether current \cite{Das:2025fav}. Therefore, even for ordinary bit threads, the relevant current should be understood as a choice of representative rather than as a uniquely determined local form. Inspired by this and \cite{Faulkner:2013ica}, we notice that the differential form of \eqref{30} is identical to \eqref{31}, in which $\boldsymbol{\mathcal{\chi}}$ is obtained by the standard calculation of Einstein gravity coupled to a scalar field. We therefore regard the formal similarity between the variation of the PEE form and the CPS surface charge form as motivation for a minimal ansatz, rather than as a derivation of an exact equivalence. More precisely, for each boundary label $x$, we ask whether there exists a generator $\tilde{\xi}$ and an associated integrable codimension-two CPS current $\tilde{\boldsymbol{j}}_{\tilde{\xi}}$ whose pullback to a spatial bulk slice represents the PEE form, possibly up to an exact form improvement. At this stage, we first formulate the un-improved candidate relation. We thus propose
\begin{equation}
\label{39}\boldsymbol{\tilde{w}}\sim\boldsymbol{\tilde{j}}_{\tilde{\xi}},\quad\boldsymbol{\tilde{k}}_{\tilde{\xi}}=\delta\boldsymbol{\tilde{j}}_{\tilde{\xi}}
\end{equation}
and
\begin{equation}
\begin{aligned}
\label{40}&\tilde{\boldsymbol{k}}_{\tilde{\xi}}=\frac{1}{2(d-1)!}\sqrt{-g}\varepsilon_{a_1\dots a_{d-1}ab}\tilde{k}^{ab} \mathrm{d}x^{a_1}\wedge\dots\wedge \mathrm{d}x^{a_{d-1}},\\
&\tilde{\boldsymbol{j}}_{\tilde{\xi}}=\frac{1}{2(d-1)!}\sqrt{-g}\varepsilon_{a_1\dots a_{d-1}ab}\tilde{j}^{ab} \mathrm{d}x^{a_1}\wedge\dots\wedge \mathrm{d}x^{a_{d-1}},\\
\end{aligned}
\end{equation}
where we have used tilde above all our quantities to distinguish them in bit threads. 
In general, the generator $\tilde{\xi}$ and its integration constants may depend parametrically on the boundary label $x$. In the explicit AdS$_3$ calculation below, however, we restrict the PEE source to the center of the boundary interval $x=0$. The label is therefore fixed throughout the matching calculation. Our test concerns this reflection-symmetric central member of the PEE flow family, rather than the complete family with arbitrary $x$. Further more, we assume that the PEE form is represented by the pullback of a current constructed from the CPS formalism
\begin{equation}
\label{33}\tilde{\boldsymbol{j}}_{\Sigma}=\boldsymbol{\tilde{j}}_{\tilde{\xi}}|_{\Sigma},\quad\boldsymbol{\tilde{k}}_{\Sigma}=\delta\boldsymbol{\tilde{j}}_{\Sigma}
\end{equation}
with $\tilde{\boldsymbol{j}}_{\Sigma}$ is pull bcak of $\tilde{\boldsymbol{j}}_{\tilde{\xi}}$ and $\tilde{\xi}$ an exact Killing vector of the Rindler-AdS$_3$ background. We then examine whether any element of the full Killing algebra can reproduce the known PEE flow after transforming back to Poincaré-AdS$_3$.
In fact, due to the ambiguity of boundary terms in CPS, a more stable way of writing is 
\begin{equation}
\boldsymbol{\tilde{w}}=\boldsymbol{\tilde{j}}_{{\Sigma}}+\mathrm{d}\boldsymbol{Y}.
\end{equation}
For the CPS candidate to represent a PEE thread form, it must satisfy the matching condition
\begin{equation}
\label{35}\int_A\boldsymbol{\tilde{w}}=\int_A\boldsymbol{\tilde{j}}_{\Sigma}+\int_A\mathrm{d}\boldsymbol{Y}=\int_A\boldsymbol{\tilde{j}}_{\Sigma}+\int_{\partial A}\boldsymbol{Y}=s_A(x).
\end{equation}
If we set $\boldsymbol{Y}|_{\partial A}=0$, we could have 
\begin{equation}
\label{36}\int_A\boldsymbol{\tilde{w}}=\int_A\boldsymbol{\tilde{j}}_{\Sigma}=s_A(x).
\end{equation}
In the remainder of this section, we set $\boldsymbol{Y}|_{\partial A}=0$ and test a fixed representative. The PEE threads flow is given by
\begin{equation}
\label{48}V^a_{cps,x}=g^{ab}(\star\tilde{\boldsymbol{j}}_{\Sigma})_b, 
\end{equation}
\subsection{A minimal Killing-generated realization
 test in Rindler-AdS$_3$}
We now test whether the central PEE thread flow, corresponding to the boundary label $r_0=0$, can be reproduced by the Killing-generated CPS candidate within the pointwise fixed generator prescription. Since the boundary label has been fixed once and for all, the Killing coefficients appearing below are treated as ordinary constants. For a general source point $r_0$, these coefficients could in principle depend parametrically on $r_0$, but such a generalization is not considered in the present work.
Since different AdS radius $L$ corresponds to different solution spaces, to simplify our problem, we consider the Rindler-AdS$_3$ situation. We first use the Killing equation in Rindler-AdS$_3$ and our codimension-2 current $\tilde{\boldsymbol{j}}_{\Sigma}$ to calculate components of PEE thread flow and then transform it back to the Poincaré AdS$_3$ by using coordinate transformations. In order to calculate the variation of the metric for convenience, the metric is given by
\begin{equation}
    ds^2=-Fd\tau^2+F^{-1}\mathrm{d}\rho^2+\rho^2 \mathrm{d}u^2,
\end{equation}
where $a$ is the Rindler acceleration parameter \cite{Das:2025fav}, $\tau$ is the Rindler time and $\rho= aL$ is the horizon and $F=\frac{\rho^2}{L^2}-a^2$. To obtain the explicit expression of $\tilde{k}^{\mu\nu}$, we can write the parametric variation as
\begin{equation}
    \label{34}\delta \hat{g}_{\mu\nu}(a)=\frac{\partial \hat{g}_{\mu\nu}}{\partial a}\delta a.
\end{equation}
The ``$\hat{g}$'' here refers to the entire spacetime metric. In the following calculation, we adopt a pointwise fixed-generator prescription. At each value of the parameter $a$, we substitute a Killing vector of the corresponding Rindler-AdS$_3$ background into the standard Iyer–Wald surface-charge form. When evaluating the phase-space variation, however, the generator is held fixed, as assumed in the derivation of the standard formula. We therefore do not include the additional contribution associated with the parametric dependence $\delta\tilde{\xi}$. Accordingly, the current obtained by integrating $\boldsymbol{\tilde{k}}_{\tilde{\xi}}$ along the $a$-direction is a candidate current within this restricted prescription, rather than the complete charge of a field-dependent symmetry.
Next, we could express the PEE thread vector field from the differential form $\boldsymbol{\tilde{j}}_{\xi}$ in differential equations system about the generator vector field $\tilde{\xi}^a$. First, use the expression \cite{Hajian:2015xlp}
\begin{equation}
\tilde{k}^{\mu\nu}=\frac{1}{16\pi G_N}[(\tilde{\xi}^{\nu}\nabla^{\mu}h-\tilde{\xi}^{\nu}\nabla_{\sigma}h^{\mu\sigma}+\tilde{\xi}_{\sigma}\nabla^{\nu}h^{\mu\sigma}+\frac{1}{2}h\nabla^{\nu}\tilde{\xi}^{\mu}-h^{\delta\nu}\nabla_{\delta}\tilde{\xi}^{\mu})-(\mu\longleftrightarrow\nu)],
\end{equation}
where $h^{\mu\nu}=\hat{g}^{\mu\sigma}\hat{g}^{\nu\gamma}\delta \hat{g}_{\sigma\gamma}$ is the metric perturbation and $h=h^{\mu}_{\mu}=\hat{g}_{\mu\nu}h^{\mu\nu}$is its trace. 
It's easy to check 
\begin{equation}
h^{\tau\tau}=\frac{2a\delta a}{F^2},\quad h^{\rho\rho}=2a\delta a,\quad h^{uu}=0,\quad h=0
\end{equation}
and  the non-vanishing derivatives of $h_{\mu\nu}$ are
\begin{equation}
\begin{aligned}
&\nabla^{\rho}h^{\tau\tau}=-\frac{4a\delta a\rho}{L^2F^2},\quad\nabla^{\rho}h^{\rho\rho}=-\frac{4a\delta a\rho}{L^2},\\
&\nabla^{\tau}h^{\tau\rho}=-\frac{4a\delta a\rho}{L^2F^2},\quad\nabla^uh^{\rho u}=\frac{2a\delta a}{\rho^3}.
\end{aligned}
\end{equation}
Then we can find that the non vanish components of $\boldsymbol{\tilde{k}}_{\tilde{\xi}}$ is
\begin{equation}
\begin{aligned}
\label{42}&\tilde{k}^{\tau\rho}_{\tilde{\xi}}=\frac{a\delta a}{8\pi G_N}(\frac{\tilde{\xi}^{\tau}}{\rho}-\partial_{\rho}\tilde{\xi}^{\tau}+\frac{1}{F^2}\partial_{\tau}\tilde{\xi}^{\rho}),\\
&\tilde{k}^{\tau u}_{\tilde{\xi}}=\frac{a\delta a}{8\pi G_N}\frac{1}{F^2}\partial_{\tau}\tilde{\xi}^{u},\\
&\tilde{k}^{\rho u}_{\tilde{\xi}}=\frac{a\delta a}{8\pi G_N}(\partial_{\rho}\tilde{\xi}^{u}+\frac{\tilde{\xi}^u}{\rho}).
\end{aligned}
\end{equation}
In three dimension's co-dimension-2 current $\boldsymbol{\tilde{k}}_{\tilde{\xi}}$ is a one form
\begin{equation}
\boldsymbol{\tilde{k}}_{\tilde{\xi}}=\frac{\sqrt{-g}}{2}\epsilon_{\mu\nu\sigma}\tilde{k}_{\tilde{\xi}}^{\mu\nu}\mathrm{d}x^{\sigma}=\rho(\tilde{k}_{\tilde{\xi}}^{\rho u}\mathrm{d}\tau-\tilde{k}_{\tilde{\xi}}^{\tau u}\mathrm{d}\rho+\tilde{k}_{\tilde{\xi}}^{\tau\rho}\mathrm{d}u),
\end{equation}
pulling back to $\tau=0$ slice we have
\begin{equation}
\label{49}\boldsymbol{\tilde{k}}_{\Sigma}=\rho(-\tilde{k}_{\tilde{\xi}}^{\tau u}\mathrm{d}\rho+\tilde{k}_{\tilde{\xi}}^{\tau\rho}\mathrm{d}u)=\rho\frac{a\delta a}{8\pi G_N}[-\frac{1}{F^2}\partial_{\tau}\tilde{\xi}^{u}\mathrm{d}\rho+(\frac{\tilde{\xi}^{\tau}}{\rho}-\partial_{\rho}\tilde{\xi}^{\tau}+\frac{1}{F^2}\partial_{\tau}\tilde{\xi}^{\rho})\mathrm{d}u].
\end{equation}
Using \eqref{33} and
\begin{equation}
\label{41}\boldsymbol{\tilde{j}}_{\Sigma}=\tilde{j}_{\rho}\mathrm{d}\rho+\tilde{j}_u\mathrm{d}u.
\end{equation}
we have
\begin{equation}
\tilde{j}_{\rho}=\int\tilde{k}_{\rho}\mathrm{d}a,\quad\tilde{j}_u=\int\tilde{k}_{u}\mathrm{d}a
\end{equation}
Since we assume that $\tilde{\xi}$ is a Killing vector in Rindler-AdS$_3$, it should additionally satisfies the following equations
\begin{equation}
\begin{aligned}
&\label{51}\partial_{\tau}\tilde{\xi}^{\rho}-F^2\partial_{\rho}\tilde{\xi}^{\tau}=0,\quad\partial_{\tau}\tilde{\xi}^{\tau}+\frac{\rho}{L^2F}\tilde{\xi}^{\rho}=0;\\
&\partial_{\rho}\tilde{\xi}^u+\frac{1}{\rho^2F}\partial_u\tilde{\xi}^\rho=0,\quad\partial_{\rho}\tilde{\xi}^{\rho}-\frac{\rho}{L^2F}\tilde{\xi}^{\rho}=0;\\
&\partial_{\tau}\tilde{\xi}^u-\frac{F}{\rho^2}\partial_u\tilde{\xi}^{\tau}=0,\quad\partial_u\tilde{\xi}^u+\frac{\tilde{\xi}^{\rho}}{\rho}=0.\\
\end{aligned}
\end{equation}We solve the above equation system and combine \eqref{42} we can further simplify \eqref{49} as
\begin{equation}
\begin{aligned}
\label{47}&\tilde{k}^{\tau\rho}_{\tilde{\xi}}=\frac{a\delta a}{8\pi G_N}(\frac{A}{\rho}-\frac{\partial_{\tau}f}{a^2\sqrt{F}}),\\
&\tilde{k}^{\tau u}_{\tilde{\xi}}=-\frac{\delta a}{8\pi G_N}\frac{\partial_{\tau}\partial_uf}{a\rho F^{3/2}},
\end{aligned}
\end{equation}
in which $A$ is constant and $f$ is a function independent of $\rho$ which can be written as
\begin{equation}
\label{45}f(\tau,u)=\sum_{\sigma=\pm 1}\sum_{\epsilon=\pm 1}c_{\sigma\epsilon}e^{(\sigma\frac{a}{L}\tau+\epsilon au)},
\end{equation}
where $c_{\sigma\epsilon}$ are undetermined coefficients (more detailed about solutions \eqref{51}, \eqref{45} can be found in \eqref{solution}). We consider calculation on $\tau=0$ surface, thus we have
\begin{equation}
\begin{aligned}
\label{46}&\partial_{\tau}f|_{\tau=0}=C\frac{a}{L}e^{au}+D\frac{a}{L}e^{-au},\\
&\partial_{\tau}\partial_uf|_{\tau=0}=C\frac{a^2}{L}e^{au}-D\frac{a^2}{L}e^{-au},
\end{aligned}
\end{equation}
where $C=c_{11}-c_{-11},\;D=(c_{1-1}-c_{-1-1})$.
Then we can further simplify equation \eqref{47} to have
\begin{equation}
\begin{aligned}
\label{50}&\tilde{k}^{\tau\rho}_{\tilde{\xi}}=(\frac{aA}{8\pi\rho G_N}-\frac{Ce^{au}+De^{-au}}{8\pi G_NL\sqrt{F}})\delta a,\\
&\tilde{k}^{\tau u}_{\tilde{\xi}}=-\frac{a(Ce^{au}-De^{-au})}{8\pi\rho G_NLF^{3/2}}\delta a.
\end{aligned}
\end{equation}
Set $a=1,$ using \eqref{48},\eqref{41},\eqref{50} we can get
\begin{equation}
\begin{aligned}
\label{44}&V^{\rho}_{cps,x}=\frac{\sqrt{\rho^2-L^2}}{16\pi G_NL^2}\sum_{n=0}^{\infty}\frac{C+(-1)^nD}{n!}\left(\frac{\rho u}{L}\right)^n\boldsymbol{B}_{L^2/\rho^2}\left(\frac{n+1}{2},\frac{1}{2}\right)-\frac{A(\sqrt{\rho^2-L^2)}}{16\pi\rho G_NL},\\
&V^u_{cps,x}=\frac{(Ce^{u}-De^{-u})}{8\pi\rho G_NL}-\frac{(C-D)(\sqrt{\rho^2-L^2})}{8\pi G_NL\rho^2}\\
&-\frac{\sqrt{\rho^2-L^2}}{16\pi G_N\rho L^2}\sum_{n=0}^{\infty}\frac{C+(-1)^nD}{n!}\left(\frac{\rho}{L}\right)^nu^{n+1}\boldsymbol{B}_{L^2/\rho^2}\left(\frac{n+1}{2},\frac{1}{2}\right),
\end{aligned}
\end{equation}
where $\boldsymbol{B}_q(p,r)=\int_0^qt^{p-1}(1-t)^{r-1}dt$ is the incomplete beta function (more detailed process about this equation can be found in \eqref{solution}). Equation \eqref{44} gives the finite candidate flow obtained by integrating the surface charge form over $a$, with the full $a$-dependence of the Killing solution included. Using the coordinate relations, we can translate \eqref{44} to Poincaré coordinates \cite{Das:2025fav} as follow: 
\begin{equation}
\begin{aligned}
\label{52}V^z_{cps,x}&=V^u_{cps,x}\frac{\partial{z}}{\partial{u}}+V^{\rho}_{cps,x}\frac{\partial{z}}{\partial{\rho}}\\
&=-\frac{rz}{L}V^u_{cps,x}(z,r)-\left(\frac{2z^2}{\sqrt{(L^2+r^2+z^2)^2-4L^2r^2}}\right)\left(\frac{L^2-r^2+z^2}{L^2-r^2-z^2}\right)V^{\rho}_{cps,x}(z,r),\\
V^r_{cps,x}&=V^u_{cps,x}\frac{\partial{r}}{\partial{u}}+V^{\rho}_{cps,x}\frac{\partial{r}}{\partial{\rho}}\\
&=\frac{L^2-r^2+z^2}{2L}V^u_{cps,x}(z,r)-\left(\frac{4z^2}{\sqrt{(L^2+r^2+z^2)^2-4L^2r^2}}\right)\left(\frac{zr}{L^2-r^2-z^2}\right)V^{\rho}_{cps,x}(z,r).
\end{aligned}
\end{equation}
We now compare the flow in \eqref{52} with the known PEE flow in \eqref{37}, restoring the AdS radius $L$ and restricting the source point to the center of the boundary interval, $r_0=0$. The candidate flow in \eqref{52} is obtained from the finite current defined by integrating the Iyer–Wald surface charge form over $a$, with the full $a$-dependence of the corresponding Killing solution included. Direct comparison shows that no choice of the Killing parameters reproduces both components of the known central PEE flow throughout the relevant bulk region.\par
This mismatch shows that the direct exact Killing vector CPS construction considered here is not sufficient to reproduce the central PEE flow. It should not be interpreted as a general obstruction to a Killing vector or covariant phase space description of PEE threads. In particular, the PEE form may differ from the candidate current by an allowed exact form term $\mathrm{d}\boldsymbol{Y}$, or a different CPS current may be required. More general choices of generators may also provide alternative realizations. The present calculation should therefore be regarded as a test of this specific construction rather than as a general no-go result.

\section{Discussion and conclusions}\label{DC}
In this work, we introduced a differential form description of PEE thread flows. Given a divergenceless PEE flow $V^a_x$, the corresponding form $\boldsymbol{\tilde{\omega}}$ is defined through its Hodge dual on a spatial bulk slice. Its closedness follows from the divergenceless condition, while its flux through a bulk surface homologous to the boundary region is required to reproduce the corresponding entanglement contour. For an interval in vacuum AdS$_3$/CFT$_2$, substituting the known PEE thread flow into this construction reproduces the standard entanglement contour. This result provides a consistency check of the differential form description for a known PEE flow.

The differential form description also gives a pointwise decomposition of perturbative bit thread. Assuming that the relation between PEE threads and bit threads continues to hold at linear order, the perturbation of a bit thread flow can be written as an integral of the perturbations of the PEE flows. This suggests that PEE forms may provide a more detailed organization of the geometric information contained in perturbative bit thread constructions. Whether this information is sufficient for a unique reconstruction of the bulk metric requires further study.

Motivated by the covariant phase space construction of ordinary bit threads, we further considered whether the PEE form can be related to a current obtained from the Iyer–Wald surface charge form
\begin{equation}
\boldsymbol{\tilde{w}}\sim\boldsymbol{\tilde{j}}_{\tilde{\xi}}|_{\Sigma}
\end{equation}
where, more generally, the relation may hold up to an exact form improvement,
\begin{equation}
\boldsymbol{\tilde{w}}=\boldsymbol{\tilde{j}}_{{\Sigma}}+\mathrm{d}\boldsymbol{Y}. 
\end{equation}
This relation should be regarded as a possible CPS description of the PEE form. We tested this proposal in Rindler-AdS$_3$. The parameter $a$ was used to parametrize a one-parameter family of backgrounds, at each value of $a$, an exact Killing vector $\tilde{\xi}(a)$ of the corresponding background was substituted into the Iyer–Wald surface charge form. Since the Killing solution itself depends on $a$, this dependence is retained in the resulting surface charge form. The finite candidate current was then defined by integrating the surface charge form over $a$. Using this construction, we solved the Killing equations in Rindler-AdS$_3$, evaluated the surface charge form, and performed the integration over $a$. The resulting one form was converted into a candidate flow on the static slice. After transforming this flow to Poincaré-AdS$_3$, we found that no choice of the Killing parameters reproduces both components of the known PEE flow in the relevant bulk region. In the explicit calculation, the PEE source was fixed at the center of the boundary interval, $r_0=0$. The present result therefore applies only to the reflection-symmetric central PEE flow.\par
The mismatch should be interpreted within the scope of this construction. It shows that the finite current obtained by integrating the exact Killing surface charge form over $a$ does not reproduce the known central PEE flow. Since the full $a$-dependence of the Killing solution is already included in the surface charge form before the integration, the mismatch should not be attributed simply to the $a$-dependence of the Killing vector. The result also does not provide a general no-go statement for a CPS description of PEE threads. One possible extension is the exact form term $\mathrm{d}\boldsymbol{Y}$. Such a term can modify the local form while preserving the required flux under suitable boundary conditions. It is therefore useful to determine whether an allowed exact form improvement can remove the mismatch. The form of $\boldsymbol{Y}$ should follow from the allowed ambiguity of the CPS construction rather than being introduced as an arbitrary fitting term. Another possibility is that the CPS current used for ordinary bit threads is too restrictive for PEE threads. PEE threads resolve the flow with respect to individual boundary points, and their local structure may require a different current or a modified relation between the current and the PEE form. A more general class of generators may also be relevant. In particular, one may consider generators beyond exact Killing vectors if the corresponding surface charge form satisfies the required closedness and flux conditions.
\par
Several questions remain open. The role of the exact form term $d\boldsymbol{Y}$, including its allowed form and boundary conditions, should be studied in detail. It would also be useful to determine whether a different CPS current can provide a more direct description of the pointwise PEE structure. The present analysis is restricted to a static interval in AdS$_3$ and to the central source $r_0=0$; extensions to arbitrary source points, higher dimensions, more general entangling regions, and time-dependent configurations would provide further tests. Finally, it would be interesting to study whether the first law of the entanglement contour \cite{Han:2021ycp} can be related to the linearized gravitational equations in a way similar to the first law of holographic entanglement entropy \cite{Lashkari:2013koa,Faulkner:2013ica,Blanco:2013joa}.\par
In summary, we provide a differential form description of PEE thread flows and explore a possible connection with covariant phase space. The explicit AdS$_3$ calculation shows that the direct exact Killing vector construction based on the Iyer–Wald surface charge form does not reproduce the known central PEE flow. This result motivates the study of exact form improvements, alternative CPS currents, and more general choices of generators.

\acknowledgments
This work was supported by the
National Natural Science Foundation of China under the
Grant No. 12375049, Key Program of the Natural Science Foundation of Jiangxi Province under the Grant
No. 20232ACB201008, and the
Ganpo High-Level Innovative Talent Program. W-C.G. is supported by the National
Natural Science Foundation of China under the Grant No. 12405064 and Jiangxi Provincial Natural Science Foundation under the Grant No. 20242BCE50055.

\appendix
\section{Differential form of bit threads}\label{app}
We review the differential forms of bit threads.
The differential $(d-1)$ forms $\boldsymbol{w}$ connected to the divergenceless vector field of bit threads $V^a_A$ could be written as \cite{Agon:2020mvu}:
\begin{equation}
V_A^a=g^{ab}(\star{\boldsymbol{w}})_b,
\end{equation}
where $\star{\boldsymbol{w}}$ represents the Hodge star dual of $\boldsymbol{w}$, defined by:
\begin{equation}
(\star{\boldsymbol{w}})_b=\frac{1}{(d-1)!}\sqrt{g}w^{\mu_1\dots\mu_{d-1}}\varepsilon_{\mu_1\dots\mu_{d-1}b}.
\end{equation}
In the above formula 
$\varepsilon_{\mu_1\dots\mu_{d}}$
represents the totally antisymmetric Levi-Civita symbol, with sign convention $\varepsilon_{i_1\dots i_{d-1}z}=1$. From the above, we could see that although the divergenceless vector field depends on the background metric, $\boldsymbol{w}$ can be defined independently of $g_{ab}$. The $(d-1)$-forms $\boldsymbol{w}$ is given by
\begin{equation}
\boldsymbol{w}=\frac{1}{(d-1)!}{\epsilon_{a_1\dots a_{d-1}b}}V_A^b \mathrm{d}x^{a_1}\wedge\dots\wedge \mathrm{d}x^{a_{d-1}},
\end{equation}
where $\boldsymbol{\epsilon}$ is volume form, given by:
\begin{equation}
\boldsymbol{\epsilon}=\frac{1}{d!}\epsilon_{\mu_1\dots \mu_d}\mathrm{d}x^{\mu_1}\wedge\dots\wedge \mathrm{d}x^{\mu_d},
\end{equation}
and $\epsilon_{\mu_1\dots\mu_d}$ is components such that $\epsilon_{i_1\dots i_{d-1}z}=\sqrt{g}$. Explicitly, taking the exterior derivative of the equation leads to
\begin{equation}
\mathrm{d}\boldsymbol{w}=(\nabla_aV^a)\boldsymbol{\epsilon}.
\end{equation}
We consider a codimension-one surface $\Gamma$, with local unit normal $n$ in terms of the $(d-1)$-form $\boldsymbol{\tilde{\epsilon}}$, thus
\begin{equation}
\boldsymbol{w|_{\Gamma}}=(n_aV^a)\boldsymbol{\tilde{\epsilon}},
\end{equation}
where $\boldsymbol{\tilde{\epsilon}}$ is induced $(d-1)$ volume by $\boldsymbol{\epsilon}$. In a bulk region $N$ with $\partial N=A\cup(-m)$, where $m$ is a surface homologous to $A\ (\text{denoted as}\ m\sim A)$, using Gauss's theorem and the divergenceless of bit thread flow, we can get 
\begin{equation}
\int_N\nabla_aV^a\boldsymbol{\epsilon}=\int_{\partial N}(n_aV^a)\boldsymbol{\tilde{\epsilon}}=\int_An_aV^a\boldsymbol{\tilde{\epsilon}}-\int_mn_aV^a\boldsymbol{\tilde{\epsilon}}=0.
\end{equation}
Now we can rewrite the above formula in differential forms $\boldsymbol{w}$:
\begin{equation}
\int_N\mathrm{d}\boldsymbol{w}=\int_{\partial N}\boldsymbol{w}=\int_A\boldsymbol{w}-\int_m\boldsymbol{w}=0.
\end{equation}
Then, translate the max flow-min cut theorem to the language of differential forms
\begin{equation}
\int_m\boldsymbol{w}=\int_m(n_aV_A^a)\tilde{\boldsymbol{\epsilon}}\le\frac{1}{4G_N}\int_m\tilde{\boldsymbol{\epsilon}},
\end{equation}
here we use $|V^a_A|\le\frac{1}{4G_N}$ which in terms of forms
can be rewritten as
\begin{equation}
\label{28}\frac{1}{(d-1)!}g^{a_1b_1}\dots g^{a_{d-1}b_{d-1}}w_{a_1\dots a_{d-1}}w_{b_1\dots b_{d-1}}\le(\frac{1}{4G_N})^2.
\end{equation}
Then, the max flow-min cut theorem then implies that
\begin{equation}
\mathop{\text{max}}\limits_{\boldsymbol{w}\in\boldsymbol{W}}\int_A\boldsymbol{w}=\frac{1}{4G_N}\mathop{\text{min}}\limits_{m\sim A}\int_m\tilde{\boldsymbol{\epsilon}},
\end{equation}
where $\boldsymbol{W}$ is the set of closed forms obeying the bound \eqref{28}.
Finally, using the ingredients described above, and combining with the RT formula for entanglement entropy, we could translate the RT formula to the language of differential forms:
\begin{equation}
S_A=\mathop{\text{max}}\limits_{\boldsymbol{w}\in\boldsymbol{W}}\int_A\boldsymbol{w}.
\end{equation}
\section{Covariant phase space formalism: a brief review}\label{CPS}
We consider a spacetime manifold of $d+1$ dimension $M$. In covariant phase space formalism \cite{Wald:1993nt,Iyer:1994ys}, Lagrangian is a $d+1$-form $\boldsymbol{L}$ in $M$ and its' first variation has the universal form
\begin{equation}
\begin{aligned}
\delta\boldsymbol{L}=\boldsymbol{E}^{\phi}\delta\phi+\mathrm{d}\boldsymbol{\Theta}(\phi,\delta\phi),
\end{aligned}
\end{equation}
where $\phi$ are all dynamical fields collectively and $\boldsymbol{E}^{\phi}=0$ are the equations of motion for the theory, $\boldsymbol{\Theta}$ is a $d$-form in $M$ and a 1-form on the phase space, which is called the symplectic potential. Then we can define the symplectic current, which is a $d$-form in $M$ and a $2$-form on the phase space as
\begin{equation}
\begin{aligned}
\boldsymbol{\omega}(\phi,\delta_1\phi,\delta_2\phi)=\delta_1\boldsymbol{\Theta}
\end{aligned}(\phi,\delta_2\phi)-\delta_2\boldsymbol{\Theta}(\phi,\delta_1\phi).
\end{equation}
Choosing a codimension-1 spacelike surface $\Sigma$, we can define the two form symplectic $\boldsymbol{\Omega}$
\begin{equation}
\begin{aligned}
\boldsymbol{\Omega}(\phi,\delta_1\phi,\delta_2\phi)=\int_{\Sigma}\boldsymbol{\omega}(\phi,\delta_1\phi,\delta_2\phi).
\end{aligned}
\end{equation}
Collecting all of the solutions of the equations of motion gives a manifold $\mathcal{P}$, then $\mathcal{P}$ with a symplectic form $\boldsymbol{\Omega}$ is our covariant phase space.\par
Let $\xi^a$ be any vector field on $M$ and consider the field variation
\begin{equation}
\begin{aligned}
\delta\phi=\mathcal{L}_{\xi}\phi
\end{aligned}
\end{equation}
by Cartan's identity
\begin{equation}
\begin{aligned}
\mathcal{L}_{\xi}\boldsymbol{\Lambda}=\xi\cdot \mathrm{d}\boldsymbol{\Lambda}+\mathrm{d}(\xi\cdot\boldsymbol{\Lambda}),
\end{aligned}
\end{equation}
where $\boldsymbol{\Lambda}$ is an arbitrary differential form and $\cdot$ denotes the contraction of a vector field with the first index of a differential form. Then we have
\begin{equation}
\begin{aligned}
\mathcal{L}_{\xi}\boldsymbol{L}=\xi\cdot \mathrm{d}\boldsymbol{L}+\mathrm{d}(\xi\cdot\boldsymbol{L})=\mathrm{d}(\xi\cdot\boldsymbol{L}).
\end{aligned}
\end{equation}
The second equality is due to the fact that the Lagrangian is a $d+1$-form. To each $\xi^a$ we can associate a Noether current $d$-form, defined by
\begin{equation}
\begin{aligned}
\label{j}\boldsymbol{J}=\boldsymbol{\Theta}(\phi,\mathcal{L}_{\xi}\phi)-\xi\cdot\boldsymbol{L},
\end{aligned}
\end{equation}
and take d of $\boldsymbol{J}$
\begin{equation}
\mathrm{d}\boldsymbol{J}=\mathrm{d}\boldsymbol{\Theta}(\phi,\mathcal{L}_{\xi}\phi)-\mathrm{d}(\xi\cdot\boldsymbol{L}).
\end{equation}
If we rewrite
\begin{equation}
\delta\boldsymbol{L} =\boldsymbol{E}^{\phi}\mathcal{L}_{\xi}\phi+\mathrm{d}\boldsymbol{\Theta}(\phi,\mathcal{L}_{\xi}\phi)=\mathrm{d}(\xi\cdot\boldsymbol{L}),
\end{equation}
therefore
\begin{equation}
\mathrm{d}\boldsymbol{J}=-\boldsymbol{E}^{\phi}\mathcal{L}_{\xi}\phi.
\end{equation}
Hence, if the equation of motion holds, then
\begin{equation}
\mathrm{d}\boldsymbol{J}=0.
\end{equation}
Then there exists a $(d-1)$-form $\boldsymbol{Q}$, locally constructed from $\xi^a$ and the fields, called the Noether charge, such that
\begin{equation}
\boldsymbol{J}=\mathrm{d}\boldsymbol{Q}.
\end{equation}
Varying equation \eqref{j} meanwhile holding $\xi^a$ fixed, then
\begin{equation}
\delta \boldsymbol{J}=\delta\boldsymbol{\Theta}(\phi,\mathcal{L}_{\xi}\phi)-\xi\cdot\delta\boldsymbol{L}.
\end{equation}
Using $\delta\boldsymbol{L}=\boldsymbol{E}^{\phi}\delta\phi+\mathrm{d}\boldsymbol{\Theta}$, we have
\begin{equation}
\xi\cdot\delta\boldsymbol{L}=\xi\cdot(\boldsymbol{E}^{\phi}\delta\phi)+\xi\cdot \mathrm{d}\boldsymbol{\Theta}.
\end{equation}
If $\boldsymbol{E}^{\phi}=0$, then
\begin{equation}
\xi\cdot\delta\boldsymbol{L}=\xi\cdot \mathrm{d}\boldsymbol{\Theta}.
\end{equation}
Using Cartan identity with $\boldsymbol{\Lambda}=\boldsymbol{\Theta}$, i.e.
\begin{equation}
\mathcal{L}_{\xi}\boldsymbol{\Theta}=\xi\cdot \mathrm{d}\boldsymbol{\Theta}+\mathrm{d}(\xi\cdot\boldsymbol{\Theta}),
\end{equation}
then we have
\begin{equation}
\xi\cdot\delta\boldsymbol{L}=\xi\cdot \mathrm{d}\boldsymbol{\Theta}=\mathcal{L}_{\xi}\boldsymbol{\Theta}-\mathrm{d}(\xi\cdot\boldsymbol{\Theta}).
\end{equation}
Plugging back into $\delta\boldsymbol{J}$, thus
\begin{equation}
\delta\boldsymbol{J}=\delta\boldsymbol{\Theta}(\phi,\mathcal{L}_{\xi}\phi)-\mathcal{L}_{\xi}\boldsymbol{\Theta}(\phi,\delta\phi)+\mathrm{d}(\xi\cdot\boldsymbol{\Theta}).
\end{equation}
Since we can write
\begin{equation}
\boldsymbol{\omega}(\phi,\delta\phi,\mathcal{L}_{\xi}\phi)=\delta\boldsymbol{\Theta}(\phi,\mathcal{L}_{\xi}\phi)-\mathcal{L}_{\xi}\boldsymbol{\Theta}(\phi,\delta\phi),
\end{equation}
hence 
\begin{equation}
\delta\boldsymbol{J}=\boldsymbol{\omega}(\phi,\delta\phi,\mathcal{L}_{\xi}\phi)+\mathrm{d}(\xi\cdot\boldsymbol{\Theta}).
\end{equation}
For on-shell case, we have $\boldsymbol{J}=\mathrm{d}\boldsymbol{Q}$ and $\delta\boldsymbol{J}=\mathrm{d}(\delta\boldsymbol{Q})$, thus
\begin{equation}
\boldsymbol{\omega}(\phi,\delta\phi,\mathcal{L}_{\xi}\phi)=\mathrm{d}(\delta\boldsymbol{Q})-\mathrm{d}(\xi\cdot\boldsymbol{\Theta})=\mathrm{d}(\delta\boldsymbol{Q}-\xi\cdot\boldsymbol{\Theta}),
\end{equation}
or we can write
\begin{equation}
\label{w}\boldsymbol{\omega}(\phi,\delta\phi,\mathcal{L}_{\xi}\phi)=\mathrm{d}\boldsymbol{k}(\phi,\delta\phi),
\end{equation}
where $\boldsymbol{k}$ is the codimension-2 current for diffeomorphism invariant theories defined by
\begin{equation}
\boldsymbol{k}=\delta\boldsymbol{Q}-\xi\cdot\boldsymbol{\Theta}.
\end{equation}
The equation \eqref{w} is called the ‘fundamental identity’ of the CPS formalism.
If one can find a $d$-form $\boldsymbol{B}$ such that $\boldsymbol{\Theta}=\delta\boldsymbol{B}$, then the above equation can be rewritten as
\begin{equation}
\boldsymbol{k}=\delta\boldsymbol{j},
\end{equation}
where the codim-2 form $\boldsymbol{j}$
is given by
\begin{equation}
\boldsymbol{j}=\boldsymbol{Q}-\xi\cdot\boldsymbol{B}.
\end{equation}
This current form is the main ingredient to construct a configuration for bit threads.
\section{Solutions of Killing equations in Rindler-AdS$_3$}\label{solution}
We should solve the following partial differential equation system
\begin{equation}
\begin{aligned}
&\partial_{\tau}\tilde{\xi}^{\rho}-F^2\partial_{\rho}\tilde{\xi}^{\tau}=0,\quad\partial_{\tau}\tilde{\xi}^{\tau}+\frac{\rho}{L^2F}\tilde{\xi}^{\rho}=0;\\
&\partial_{\rho}\tilde{\xi}^u+\frac{1}{\rho^2F}\partial_u\tilde{\xi}^\rho=0,\quad\partial_{\rho}\tilde{\xi}^{\rho}-\frac{\rho}{L^2F}\tilde{\xi}^{\rho}=0;\\
&\partial_{\tau}\tilde{\xi}^u-\frac{F}{\rho^2}\partial_u\tilde{\xi}^{\tau}=0,\quad\partial_u\tilde{\xi}^u+\frac{\tilde{\xi}^{\rho}}{\rho}=0.\\
\end{aligned}
\end{equation}
From
\begin{equation}
\partial_{\rho}\tilde{\xi}^{\rho}-\frac{\rho}{L^2F}\tilde{\xi}^{\rho}=0
\end{equation}
and we have
\begin{equation}
\partial_{\rho}(\sqrt{F})=\frac{\partial_{\rho}F}{2\sqrt{F}}
\end{equation}
So we can write
\begin{equation}
\tilde{\xi}^{\rho}=\sqrt{F}f(\tau,u).
\end{equation}
From
\begin{equation}
\partial_{\tau}\tilde{\xi}^{\rho}-F^2\partial_{\rho}\tilde{\xi}^{\tau}=0
\end{equation}
and by substituting $\xi^{\rho}$ we have
\begin{equation}
\partial_{\rho}\tilde{\xi}^{\tau}=\frac{\partial_{\tau}f}{F^{3/2}}.
\end{equation}
Integrating in $\rho$, then
\begin{equation}
\label{c1}\tilde{\xi}^{\tau}=A(\tau,u)-\frac{\rho}{a^2\sqrt{F}}\partial_{\tau}f(\tau,u).
\end{equation}
Similarly, from
\begin{equation}
\partial_{\rho}\tilde{\xi}^u+\frac{1}{\rho^2F}\partial_u\tilde{\xi}^\rho=0
\end{equation}
we obtain
\begin{equation}
\label{c2}\tilde{\xi}^u=B(\tau,u)-\frac{\sqrt{F}}{a^2\rho}\partial_uf(\tau,u).
\end{equation}
Substitute \eqref{c1} into $\partial_{\tau}\tilde{\xi}^{\tau}+\frac{\rho}{L^2F}\tilde{\xi}^{\rho}=0$, thus
\begin{equation}
\partial_{\tau}\tilde{\xi}^{\tau}=\partial_{\tau}A(\tau,u)-\frac{\rho}{a^2\sqrt{F}}\partial^2_{\tau}f(\tau,u)=-\frac{\rho}{L^2\sqrt{F}}f(\tau,u).
\end{equation}
Since the first term is independent of $\rho$, while $\frac{\rho}{\sqrt{F}}$ is not, the two parts must vanish separately:
\begin{equation}
\partial_{\tau}A(\tau,u)=0,\quad\partial^2_{\tau}f(\tau,u)=\frac{a^2}{L^2}f(\tau,u).
\end{equation}
Similarly, Substituting \eqref{c2} into $\partial_u\tilde{\xi}^u+\frac{\tilde{\xi}^{\rho}}{\rho}=0$ gives
\begin{equation}
\partial_{u}B(\tau,u)=0,\quad\partial^2_uf(\tau,u)=a^2f(\tau,u). 
\end{equation}
Finally, by using $\partial_{\tau}\tilde{\xi}^u-\frac{F}{\rho^2}\partial_u\tilde{\xi}^{\tau}=0$ we have
\begin{equation}
\rho^2\partial_{\tau}B(\tau,u)=F\partial_uA(\tau,u).
\end{equation}
Because $F=\rho^2/L^2-a^2$, equality for all $\rho$ implies, for $a\ne0$,
\begin{equation}
\partial_uA(\tau,u)=\partial_{\tau}B(\tau,u)=0.
\end{equation}
Thus $A$ and $B$ are constants. So we can finally get
\begin{equation}
\begin{aligned}
&\tilde{\xi}^{\tau}=A-\frac{\rho}{a^2\sqrt{F}}\partial_{\tau}f(\tau,u),\\
&\tilde{\xi}^{\rho}=\sqrt{F}f(\tau,u),\\
&\tilde{\xi}^u=B-\frac{\sqrt{F}}{a^2\rho}\partial_uf(\tau,u)
\end{aligned}
\end{equation}
and
\begin{equation}
f(\tau,u)=\sum_{\sigma=\pm 1}\sum_{\epsilon=\pm 1}c_{\sigma\epsilon}\exp{\left(\sigma\frac{a}{L}\tau+\epsilon au\right)}.
\end{equation}
\section{Derivation of components of $V^a_{cps,x}$ in Rindler-AdS$_3$}
We have
\begin{equation}
\begin{aligned}
&\tilde{j}_{u}=\rho\int_0^1\left(\frac{aA}{8\pi\rho G_N}-\frac{Ce^{au}+De^{-au}}{8\pi G_NL\sqrt{F}}\right)\mathrm{d}a.
\end{aligned}
\end{equation}
As we set $a=1$ later, we can write the second term above as follows
\begin{equation}
\mathcal{F}=\frac{1}{8\pi G_NL}\int_0^a\frac{Ce^{ux}+De^{-ux}}{\sqrt{E^2-x^2}}\mathrm{d}x,
\end{equation}
where we define $E=\rho/L$. We can expand the exponential and integrate term by term, the required monomial integral is
\begin{equation}
\int_0^a\frac{x^n}{\sqrt{E^2-x^2}}dx=\frac{E^n}{2}\boldsymbol{B}_{a^2/E^2}\left(\frac{n+1}{2},\frac{1}{2}\right),
\end{equation}
where $\boldsymbol{B}_q(p,r)=\int_0^qt^{p-1}(1-t)^{r-1}dt$ is the incomplete beta function. It yields that
\begin{equation}
\mathcal{F}(1)=\frac{1}{16\pi G_NL}\sum_{n=0}^{\infty}\frac{C+(-1)^nD}{n!}\left(\frac{\rho u}{L}\right)^n\boldsymbol{B}_{L^2/\rho^2}\left(\frac{n+1}{2},\frac{1}{2}\right).
\end{equation}
We have
\begin{equation}
\begin{aligned}
\tilde{j}_{\rho}
&=\rho\int_0^1\frac{a(Ce^{au}-D^{-au})}{8\pi\rho G_NLF^{3/2}}\mathrm{d}a\\
&=\frac{1}{8\pi G_NL}[\frac{Ce^{au}-De^{-au}}{\sqrt{F}}-\frac{L(C-D)}{\rho}-u\int_0^1\frac{Ce^{au}+De^{-au}}{\sqrt{F}}\mathrm{d}a],
\end{aligned}
\end{equation}
where we used integration by parts in the second equal. Similar to the previous approach, we can set
\begin{equation}
\mathcal{G}=\frac{u}{8\pi G_NL}\int_0^a\frac{Ce^{ux}+De^{-ux}}{\sqrt{E^2-x^2}}\mathrm{d}x.
\end{equation}
Then we can write
\begin{equation}
\begin{aligned}
\mathcal{G}(1)=\frac{1}{16\pi G_NL}\sum_{n=0}^{\infty}\frac{C+(-1)^nD}{n!}\left(\frac{\rho}{L}\right)^nu^{n+1}\boldsymbol{B}_{L^2/\rho^2}\left(\frac{n+1}{2},\frac{1}{2}\right).
\end{aligned}
\end{equation}
Using \eqref{48}, we have
\begin{equation}
\begin{aligned}
&V^{\rho}_{cps,x}=-\frac{\sqrt{F}}{\rho}[\frac{A}{16\pi G_N}-\rho\mathcal{F}(1)]=\frac{\sqrt{\rho^2-L^2}}{16\pi G_NL^2}\sum_{n=0}^{\infty}\frac{C+(-1)^nD}{n!}\left(\frac{\rho u}{L}\right)^n\boldsymbol{B}_{L^2/\rho^2}\left(\frac{n+1}{2},\frac{1}{2}\right)-\frac{A(\sqrt{\rho^2-L^2)}}{16\pi\rho G_NL},\\
&V^{u}_{cps,x}=\frac{\sqrt{F}}{\rho}[\frac{1}{8\pi G_NL}(\frac{Ce^{u}-De^{-u}}{\sqrt{\frac{\rho^2}{L^2}-1}}-\frac{L(C-D)}{\rho})-\mathcal{G}(1)]\\
&=\frac{(Ce^{u}-De^{-u})}{8\pi\rho G_NL}-\frac{(C-D)(\sqrt{\rho^2-L^2})}{8\pi G_NL\rho^2}-\frac{\sqrt{\rho^2-L^2}}{16\pi G_N\rho L^2}\sum_{n=0}^{\infty}\frac{C+(-1)^nD}{n!}\left(\frac{\rho}{L}\right)^nu^{n+1}\boldsymbol{B}_{L^2/\rho^2}\left(\frac{n+1}{2},\frac{1}{2}\right).\\
\end{aligned}
\end{equation}

\bibliographystyle{JHEP}
\bibliography{PEE.bib}

\end{document}